\documentclass[prd,superscriptaddress,amsfonts,amssymb,amsmath,showpacs,twocolumn,nofootinbib]{revtex4-2}
\usepackage{bm}
\usepackage{amsfonts}
\usepackage{latexsym}
\usepackage{graphicx}
\usepackage{amsmath}
\usepackage{palatino}
\usepackage{xcolor} 
\usepackage{mathpazo}
\usepackage{tensor}
\usepackage{rotating}
\usepackage{textcomp}
\usepackage{float}
\usepackage{booktabs}
\usepackage{dcolumn}
\usepackage[titletoc]{appendix}
\usepackage{booktabs}
\usepackage{multirow}
\usepackage{hyperref}
\hypersetup{colorlinks,citecolor=blue}
\usepackage{amsmath}
\usepackage{xcolor}
\usepackage{orcidlink}
\usepackage{epsfig}
\usepackage{caption}
\usepackage{subcaption}
\usepackage{commath}
\hypersetup{colorlinks,citecolor=blue}
\hypersetup{colorlinks=true,linkcolor=magenta,filecolor=magenta,    urlcolor=blue}

\def\be{\begin{equation}}
\def\ee{\end{equation}}
\def\bea{\begin{eqnarray}}
\def\eea{\end{eqnarray}}

\begin{document}

\title{$\Lambda(t)$CDM Model: Cosmological Implications and Dynamical System Analysis}

\author{Himanshu Chaudhary}
\email{himanshu.chaudhary@ubbcluj.ro,\\
himanshuch1729@gmail.com}
\affiliation{Department of Physics, Babeș-Bolyai University, Kogălniceanu Street, Cluj-Napoca, 400084, Romania}
\affiliation{Research Center of Astrophysics and Cosmology, Khazar University, Baku, AZ1096, 41 Mehseti Street, Azerbaijan}
\author{Ratul Mandal}
\email{ratulmandal2022@gmail.com} 
\affiliation{Department of
Mathematics, Indian Institute of Engineering Science and Technology, Shibpur, Howrah-711 103, India}
\author{Masroor Bashir}
\email{masroor.bashir@iiap.res.in} 
\affiliation{Indian Institute of Astrophysics, Koramangala II Block, Bangalore 560034, India}
\affiliation{Department of Physics, Pondicherry University, R.V. Nagar, Kalapet, 605 014, Puducherry, India}
\author{Vipin Kumar Sharma}
\email{vipinkumar.sharma@iiap.res.in\\vipinastrophysics@gmail.com}
\affiliation{Indian Institute of Astrophysics, Koramangala II Block, Bangalore 560034, India}
\author{Ujjal Debnath}
\email{ujjaldebnath@gmail.com} 
\affiliation{Department of Mathematics, Indian Institute of Engineering Science and Technology, Shibpur, Howrah-711 103, India}

\begin{abstract}
We investigated a time-varying cosmological constant model using recent BAO measurements from DESI DR2, combined with Type Ia supernova samples (Pantheon$^{+}$, DES-Dovekie, and Union3) and CMB shift parameters, to constrain the $\Lambda(t)$CDM model parameters via Markov Chain Monte Carlo analysis. We find that the interaction term $Q(z)$ shows a sign change for all dataset combinations by crossing $Q(z)=0$, depending on the choice of the dataset: at low redshift $Q(z)<0$, indicating vacuum energy decaying into dark matter, while at high redshift $Q(z)>0$, corresponding to dark matter decaying into vacuum energy. The dynamical system analysis found three critical points, namely $P_1,P_2$, and $P_3$ respectively. The resulting critical points, determined by the underlying cosmological parameters, correspond to distinct epochs in cosmic evolution. Depending on the parameter combinations, these points characterize various cosmological phases, ranging from an accelerated stiff matter-dominated era to late-time accelerated expansion. The stability of each critical point is analyzed using linear stability theory, with the relevant physical constraints on the cosmological parameters duly incorporated throughout the analysis. For each dataset combinations, the $\Lambda(t)$CDM model predicts that $\omega_0 > -1$, showing a preference for dynamical dark energy over the cosmological constant scenario with $\omega_0 = -1$. Consequently, the model exhibits a transition phase in the range $N \equiv \log a(t) \approx -0.51$ to $-0.48$ and predicts $q_0$ in the range $-0.54$ to $-0.52$, with the precise transition point depending on the choice of dataset. Finally, the Bayesian evidence shows strong support for the $\Lambda(t)$CDM model over $\Lambda$CDM
\end{abstract}

\maketitle

\tableofcontents
\section{Introduction}\label{sec_1}
The $\Lambda$CDM model commonly known as the concordance model of cosmology has achieved remarkable success in describing the statistical properties and morphology of large-scale structure (LSS) and evolution of the universe. It postulates a specially flat universe dominated by cold dark matter (CDM) and a positive cosmological constant ($\Lambda$), the latter being responsible for observed late-time acceleration \cite{Aghanim:2018eyx,Planck:2018vyg,Riess_1998,Padmanabhan:2002ji,Copeland:2006wr,Sahni:1999gb}. Despite its empirical robustness, the $\Lambda$CDM framework is not without theoretical and observational tensions  \cite{Park:2017xbl,Perivolaropoulos:2021jda,Bullock:2017xww,Hu:2023jqc}. Chief among these is the cosmological constant problem, which arises from the striking discrepancy between the observed value of $\Lambda$ and the vastly larger vacuum energy density predicted by quantum field theory posing a significant fine-tuning challenge \cite{Joan-2013,Tian:2019enx,SolaPeracaula:2024nsz,Diaz-Pachon:2024nsq,Jain:2014wsa,CHANKOWSKI199928}. Equally puzzling is the coincidence problem, which asks why the energy densities of dark energy and dark matter are of the approximately same order in the current epoch \cite{Perivolaropoulos:2021jda,Velten:2014nra}. Moreover, $\Lambda$CDM struggles to resolve the growing Hubble tension, as well as the $\sigma_{8}$ tension, which highlight inconsistencies between early and late universe measurements of the Hubble constant and the amplitude of matter fluctuations, respectively \cite{Bhattacharyya:2018fwb,Pandey:2019plg,Hu:2023jqc,DiValentino:2021izs,pan2025interacting,halder2024phase,li2024constraints,kritpetch2025interacting,wang2025resolving}. In response to these challenges, various theoretical frameworks have been proposed, including dynamical dark energy models  \cite{Yang:2021eud,Li:2020ybr,Sharma:2025qmv,Dinda:2024kjf,Chevallier:2000qy,Jassal:2004ej}, interacting dark energy scenarios \cite{Giare:2024smz,Li:2025owk,Clemson:2011an,Li:2014eha,Xia:2016vnp}, and modifications to general relativity  \cite{Clifton:2011jh,Amendola,Shankaranarayanan:2022wbx,Dolgov:2003px,KumarSharma:2022qdf,DeFelice:2010aj,Tsujikawa:2007xu,Sharma:2019yix,Sharma:2020vex,Sharma:2023vme}.

Motivated by the latest DESI observations (that suggest dynamical behaviour of $\Lambda$), we focus here on investigating time-dependent cosmological constant models \cite{Alcaniz:2006ay}. In this regard, the simplest examples of interacting dark matter/dark energy models are scenarios with vacuum decay commonly referred to as $\Lambda(t)$ models \cite{John:1996pd,Wang:2004cp,Carneiro:2006yv}. In these models $\Lambda$ evolves as a function of cosmic time, potentially alleviating some of the theoretical inconsistencies associated with constant $\Lambda$. Such models have also been explored as viable alternatives capable of explaining recent tensions in cosmological data, particularly discrepancies in Hubble constant ($H_0$) and amplitude of matter fluctuation ($\sigma_{8}$) inferred from early and late time studies \cite{Alcaniz:2005dg,Macedo:2023zrd}.

Our work explores the cosmological implications of a specific class of $\Lambda(t)$CDM models in which the cosmological constant term evolves as the time-dependent cosmological constant term:
\begin{equation}
\Lambda(t) = \alpha a^{-2} + \beta H^{2} + \lambda_\ast,
\end{equation}
with $\alpha, \beta$ as constant parameters.
This form is physically motivated by the combination of phenomenological arguments and theoretical considerations that have appeared in the literature \citep{PhysRevD.46.2404, Macedo:2023zrd}. Each contribution can be associated with a distinct physical origin:
\begin{itemize}
    \item \textbf{The $a^{-2}$ term ($\alpha a^{-2}$):}  
    This component mimics the behavior of a curvature term in the Friedmann equations. Dimensional arguments from quantum cosmology suggest that $\Lambda \propto a^{-m}$ with $m=2$ is natural, since it decays in the same way as the curvature term  \cite{Macedo:2023zrd}. Historically, this form was proposed by Özer and Taha \cite{Ozer1987,Ozer1986} as a mechanism to alleviate cosmological problems such as the horizon and flatness issues.
    \item \textbf{The $H^{2}$ term ($\beta H^{2}$):}  
    This contribution arises naturally in the context of the Running Vacuum Model (RVM), which is motivated by renormalization group arguments in quantum field theory in curved spacetime \cite{SolaPeracaula:2021gxi,SolaPeracaula:2023swx}. In this framework, the vacuum energy density evolves as a series expansion in powers of the Hubble parameter and its derivatives, with the leading correction beyond the constant term proportional to $H^{2}$ . Physically, this reflects the idea that vacuum energy is not strictly constant but ``runs'' with the energy scale of the universe, here identified with $H$.
    \item \textbf{The constant term ($\lambda_\ast$):}  
    This corresponds to the usual cosmological constant of the $\Lambda$CDM model, ensuring that the parameterization reduces to the concordance scenario in the appropriate limit. As shown in \cite{Macedo:2023zrd}, observational constraints disfavour models with only time-variable terms, thereby requiring the presence of a constant contribution.
\end{itemize}
The combination of these three terms provides a flexible framework that interpolates between different theoretical motivations. The $\alpha a^{-2}$ term captures curvature-like decay behavior, the $\beta H^{2}$ term incorporates quantum field theoretical running of vacuum energy, and the constant $\lambda_\ast$ ensures consistency with late-time observations. As emphasized in \cite{Macedo:2023zrd}, this hybrid form is phenomenologically rich, observationally testable, and capable of addressing both theoretical puzzles (cosmological constant and coincidence problems) and observational tensions (e.g., the Hubble tension). Here in this work we have updated the constraints with recent DESI DR2 datasets. These terms are inspired by different theoretical considerations, such as quantum field theory in curved space and vacuum decay models \cite{John:1996pd,Wang:2004cp,Carneiro:2006yv}.

In this paper, we go a little further in our investigation and study new observational consequences of the $\Lambda(t)$CDM scenario with recent data sets and using a dynamical systems approach. This enables a systematic analysis of the model's background evolution, stability properties, and critical points. Further, such investigation allows us to understand the qualitative behavior of the cosmological model across different epochs and to identify stable attractors corresponding to various evolutionary phases of the Universe.

The current work is presented as follows: In Section \ref{sec_2}, we provide an overview of the mathematical formulation and discuss the proposed model. Section \ref{sec_3} discusses the Methodology and recent observational data used in this work, along with the constraints obtained on the cosmological parameters. Section \ref{sec_3}, is devoted to the dynamical system approach to compare the models. Finally, we conclude with our results and findings in Section \ref{sec_5}.

\section{Theoretical Background}\label{sec_2}
We assume that the spacetime is homogeneous, isotropic and spatially flat. This is given by the Friedmann-Lemaitre-Robertson-Walker (FLRW) spacetime metric 
\begin{equation}\label{metric}
    ds^2=-dt^2+a^2(t)\left({dr^2}+r^2d\Omega^2\right),
\end{equation}
here $a(t)$ is the usual scale factor and $d\Omega^2=d\theta^2+{\sin{\theta}}^2d\phi^2$.\\
The Einstein-Hilbert action for $\Lambda$CDM model is 
\begin{equation}\label{action}
    \mathcal{S}=\frac{1}{16\pi G}\int{d^4x\sqrt{-g}\left(R-2\Lambda\right)}+\mathcal{S}_m.
\end{equation}
Here $\Lambda$ is the cosmological constant,  $\mathcal{S}_m(g_{\mu\nu}, \Psi_m )$ represents the action the action of the standard matter component with the matter field $\Psi_m $, and $g$ is the determinant of the metric tensor $g_{\mu\nu}$. $G$ is the Newtonian gravitational constant. The cosmological implications of varying $\Lambda$CDM models is discussed in \citep{John:1996pd,Wang:2004cp,Carneiro:2006yv}.

Now varying the action  (equation \eqref{action})  with respect to $g_{\mu\nu}$, we get the Einstein field equation in its usual form as
\begin{equation}\label{efe}
    R_{\mu\nu}-\frac{1}{2}R g_{\mu\nu}+\Lambda g_{\mu\nu}=8\pi G T_{\mu\nu}
\end{equation}
In the above equation $R_{\mu\nu}$ is the Ricci tensor, $R=g^{\mu\nu}R_{\mu\nu}$ is the Ricci scalar term and $T_{\mu\nu}$ is the usual energy-momentum-stress tensor of a perfect fluid 
\begin{equation}
    T_{\mu\nu}^{\text{}}=\left(\rho+p\right)u_\mu u_\nu+p g_{\mu\nu},
\end{equation}
where $\rho$ and $p$ are density and pressure, respectively, and $u_\mu$ is the usual four velocity vector that satisfies the condition $u_\mu u^\mu=-1$ and $u^\mu \nabla_\nu u_\mu=0$ respectively. One sees by Bianchi identities that when the energy-momentum tensor is conserved  $\nabla^\mu T_{\mu\nu}=0$, it follows necessarily that $\Lambda$= constant.

Here, we consider $\Lambda$ to vary with time, so the total energy-momentum tensor is modified accordingly. We write:
\begin{equation}
T_{\mu\nu}^{\text{total}} = T_{\mu\nu} + T_{\mu\nu}^{\Lambda}
\end{equation}
where the contribution from the time-dependent cosmological term is given by:
\begin{equation}
T_{\mu\nu}^{\Lambda} = -\frac{\Lambda}{8\pi G} g_{\mu\nu}
\end{equation}
This form ensures that the cosmological term acts as a dynamic vacuum energy component, influencing the evolution of the Universe through its time dependence.
The conservation equation corresponding to the matter and the dark energy sector can be obtained by considering the vanishing four divergence of the total energy-momentum tensor i.e, $\nabla^\mu T_{\mu\nu}^{\text{total}}=0$, in the following form 
\begin{eqnarray}\label{eq_89}
    \dot{\rho}_m+3H\rho_m&=&Q
    \end{eqnarray}
    \begin{eqnarray}\label{eq_90}
    \dot{\rho}_\Lambda&=&-Q
\end{eqnarray}
Here, $Q$ represents the interaction between the dust matter and the dark energy component. By considering the above assumption, we can write the Friedmann's equation in the following form 
\begin{eqnarray}
    3H^2&=&8\pi G\left(\rho_m+\rho_\Lambda\right)\label{eqn10}\\
   \dot{H}+H^2&=&-\frac{4\pi G}{3}\left(\rho_m-2\rho_\Lambda\right)\label{eqn11}
\end{eqnarray}
Now multiplying the equation \eqref{eqn11} by 2, and adding with the equation \eqref{eqn10}, we get   
\begin{equation}
    2\dot{H}+3H^2=8\pi G \rho_\Lambda
\end{equation}
Considering 
$8\pi G=1$ and $\Lambda=8\pi G \rho_\Lambda$, we get the Friedmann field equations in the form
\begin{eqnarray}
    3H^2&=&\rho_m+\Lambda\label{eq13}\\
    2\dot{H}+3H^2&=&\Lambda\label{eq14}
\end{eqnarray}
Now, from the Friedmann equations, the expression of the Hubble parameter $H$ can be obtained as a function of redshift $z$ and $\Lambda$ by considering $\frac{d}{dt}=-H(1+z)\frac{d}{dz}$ as

\begin{equation}\label{eq014}
    \frac{dE}{dz}=\frac{3E}{2\left(1+z\right)}-\frac{\Lambda}{2EH_0^2\left(1+z\right)}
\end{equation}
Here, $E\left(z\right)=\frac{H\left(z\right)}{H_0}$  and $H_0$ is the present value of the Hubble parameter. For a given $\Lambda$, \eqref{eq014} can be solved in order to obtain the Universe evolution $E(z)$. 

In this study, we consider a generalized form of $\Lambda (= \alpha^\prime a^{-2} + \beta H^{2} + \lambda_{*})$ with $\alpha$, $\beta$ and $\lambda_{*}$ (a "bare" cosmological term) are constants \citep{macedo2023cosmological}. Such generalization allows us to investigate both observational constraints and the associated dynamical system behavior. To analyze the cosmological scenario, we have the background density parameter corresponding to the matter and the dark energy sector, respectively, as
\begin{equation}\label{densityyy}
    \Omega_m=\frac{\rho_m}{3H^2},     \Omega_\Lambda=\frac{\Lambda}{3H^2}
\end{equation}
From the above relation between the density parameters, one can classify a matter-dominated cosmological solution for $\Omega_m=1$ with $\Omega_\Lambda=0$ and a complete dark energy-dominated solution as $\Omega_m=0$ and $\Omega_\Lambda=1$. We also determine the interaction term for the $\Lambda(t)$CDM model in order to analyze its behavior later. For the $\Lambda(t)$CDM model, the interaction term is expressed as
\begin{equation}
Q(z) \equiv 2\alpha^{'}(1 + z)^2 E(z) + \beta E(z)(1 + z) \frac{dE^2(z)}{dz} 
\end{equation}

The effective equation of state (EoS) $\omega_{eff}$  and the deceleration parameter $q$ are defined as
\begin{eqnarray}
    \omega_{eff}&=&-1-\frac{2\dot{H}}{3H^2}\\
    q&=&-1-\frac{\dot{H}}{H^2}
\end{eqnarray}
The various stages of cosmological evolution can be characterized according to the value of EoS parameter $\omega_{eff}$. Specifically, for $\omega_{eff}=-1$ corresponds to a De-sitter era where the dark energy component mimics the cosmological constant behavior. A value of $\omega_{eff}=0$ indicates the matter dominated era and the quintessence era is associated with $-1<\omega_{eff}<-\frac{1}{3}$. Similarly, for $q<0$, one can obtain the accelerated expansion phase, and $q>0$ represents the decelerated phase.  

\section{Methodology and Datasets}\label{sec_3}
In this study, we constrain the parameters of the $\Lambda(t)$CDM cosmological model using the \texttt{SimpleMC} cosmological inference code \cite{simplemc,aubourg2015}. The parameter estimation is performed with the Metropolis–Hastings Markov Chain Monte Carlo (MCMC) algorithm \cite{hastings1970monte}, which efficiently explores the multidimensional parameter space and provides robust posterior distributions for the model parameters. The convergence of the MCMC chains is assessed using the Gelman–Rubin diagnostic $R-1$ \cite{gelman1992inference}, and the simulations are continued until the condition $R-1 < 0.01$ is achieved, ensuring reliable sampling of the posterior distributions.

During our analysis, we choose uniform priors for each cosmological parameter: $h \in [0, 1]$, $\Omega_{m0} \in [0, 1]$, $\Omega_{b}h^2 \in [0, 0.1]$, $\alpha \in [-2, 2]$, and $\beta \in [-5, 15]$. To analyze and visualize the results, we employ the \texttt{getdist} package\footnote{\url{https://github.com/cmbant/getdist}}, which produces detailed marginalized posterior distributions and parameter correlation plots. In our analysis, we use a combination of observational datasets: Baryon Acoustic Oscillations from Dark Energy Spectroscopic Instrument Data Release 2, Type Ia Supernovae , and the CMB compressed likelihood. Below, we describe each datasets.
\begin{itemize}
\item \textbf{Baryon Acoustic Oscillation :} First, we use the baryon acoustic oscillation (BAO) measurements from more than 14 million galaxies and quasars obtained from the Dark Energy Spectroscopic Instrument (DESI) Data Release 2 (DR2) \cite{karim2025desi}. These measurements are extracted from various tracers, including BGS, LRGs (LRG1–3), ELGs (ELG1–2), QSOs, and the Lyman-$\alpha$ forest. To analyze these measurements, we compute three key distance measures: the Hubble distance $D_H(z) = \frac{c}{H(z)},$
the comoving angular diameter distance $D_M(z) = c \int_0^z \frac{dz'}{H(z')},$ and the volume-averaged distance $D_V(z) = \left[ z\, D_M^2(z)\, D_H(z) \right]^{1/3}.$
These quantities are used to form the dimensionless ratios $D_H(z)/r_d$, $D_M(z)/r_d$, and $D_V(z)/r_d$, as the BAO data are provided in these forms. Here, $r_d$ is the sound horizon at the drag epoch, defined as $r_d = \int_{z_d}^{\infty} \frac{c_s(z)}{H(z)}\, dz,$
where $c_s(z)$ is the sound speed of the photon–baryon fluid. In the flat $\Lambda$CDM model, this yields $r_d = 147.09 \pm 0.26$ Mpc \cite{aghanim2020planck}.

\item \textbf{Type Ia supernova :} We also use the unanchored supernova (SNe Ia) dataset, which includes 1,701 light curves from 1,550 Type Ia supernovae (SNe Ia) \cite{brout2022pantheon}. We consider only the SNe Ia measurements within the redshift range $0.01 \leq z \leq 2.26$, excluding those with $z < 0.01$, as such low-redshift data are affected by significant systematic uncertainties due to peculiar velocities. We also use the re-calibrated 1,820 photometric Type Ia supernova light curves obtained over five years by the Dark Energy Survey Supernova Program (DES-Dovekie)~\cite{popovic2025dark}. This catalog consists of 1,623 DES SNe~Ia covering the redshift range $0.025 < z < 1.14$, with 197 low-redshift ($z < 0.1$) SNe~Ia from the CfA3-4/CSP Foundation sample~\cite{hicken2009cfa3,hicken2012cfa4,foley2017foundation}. The revised DES-Dovekie has 1,718 SNe~Ia overlapping between DES-Dovekie and DES SN5YR~\cite{abbott2024dark}. Finally, we consider the Union3 catalog~\cite{rubin2025union}. It consists of 2,087 Type Ia supernovae, with 1,363 of them overlapping the Pantheon$^{+}$ compilation. During our analysis, we marginalize over the nuisance parameter $\mathcal{M}$; see Eqs (A9–A12) in \cite{goliath2001supernovae} for further details.

\item \textbf{CMB Compressed Likelihood:} Finally, we use the CMB compressed likelihood approach, which effectively encapsulates the main geometric information from the full CMB power spectra into a few well-defined parameters. This method allows us to include CMB constraints without the need for a full Boltzmann code evaluation at each sampling step. Specifically, we employ a three-parameter compression scheme involving the shift parameters $R$ and $\ell_a$, along with the physical baryon density $\omega_b$. These quantities are defined as $R = \sqrt{\Omega_{m} H_0^2}\, D_M(z_\ast)$ and $\ell_a = \pi\, \frac{D_M(z_\ast)}{r_s(z_\ast)}$, where $D_M(z_\ast)$ is the comoving angular diameter distance to the surface of last scattering, and $r_s(z_\ast)$ is the comoving sound horizon at recombination \cite{wang2007observational}. The CMB information is then encoded as a multivariate Gaussian likelihood in the parameter vector $\mathbf{v} = \{R, \ell_a, \omega_b\}$, characterized by the mean values and covariance matrix calibrated from the Planck 2018 data.
\end{itemize}
The posterior distributions of the parameters in the $\Lambda(t)$CDM model are obtained by maximizing the overall likelihood function, which is given by: $\mathcal{L}_{\text{tot}} = \mathcal{L}_{\text{BAO}} \times \mathcal{L}_{\text{SNe Ia}} \times \mathcal{L}_{\text{CMB}}$.

To compare the $\Lambda(t)$CDM model with the standard $\Lambda$CDM model, we use the $\ln \mathcal{Z}$ computed using the MCEvidence~\cite{heavens2017marginal}, which is integrated into the SimpleMC code. The Bayesian evidence, ($\ln Z$), provides a quantitative assessment of how well a statistical model describes the observed data. It enables direct comparison between two cosmological models, (a) and (b), through the Bayes factor $B_{ab} \equiv \frac{Z_a}{Z_b},$ or equivalently, the relative log-Bayes evidence $\ln B_{ab} \equiv \Delta \ln Z.$ The model with the smaller ($|\ln Z|$) is considered the preferred one. To evaluate the significance of the comparison, we adopt Jeffreys’ scale \cite{jeffreys1961theory}: weak evidence corresponds to ($0 \leq |\Delta \ln Z| < 1$), moderate evidence to ($1 \leq |\Delta \ln Z| < 3$), strong evidence to ($3 \leq |\Delta \ln Z| < 5$), and decisive evidence to ($|\Delta \ln Z| \geq 5$), in favor of the model with the higher Bayesian support.

\subsection{Observational and Statistical Results}
\begin{figure*}
\begin{subfigure}{.82\textwidth}
\includegraphics[width=\linewidth]{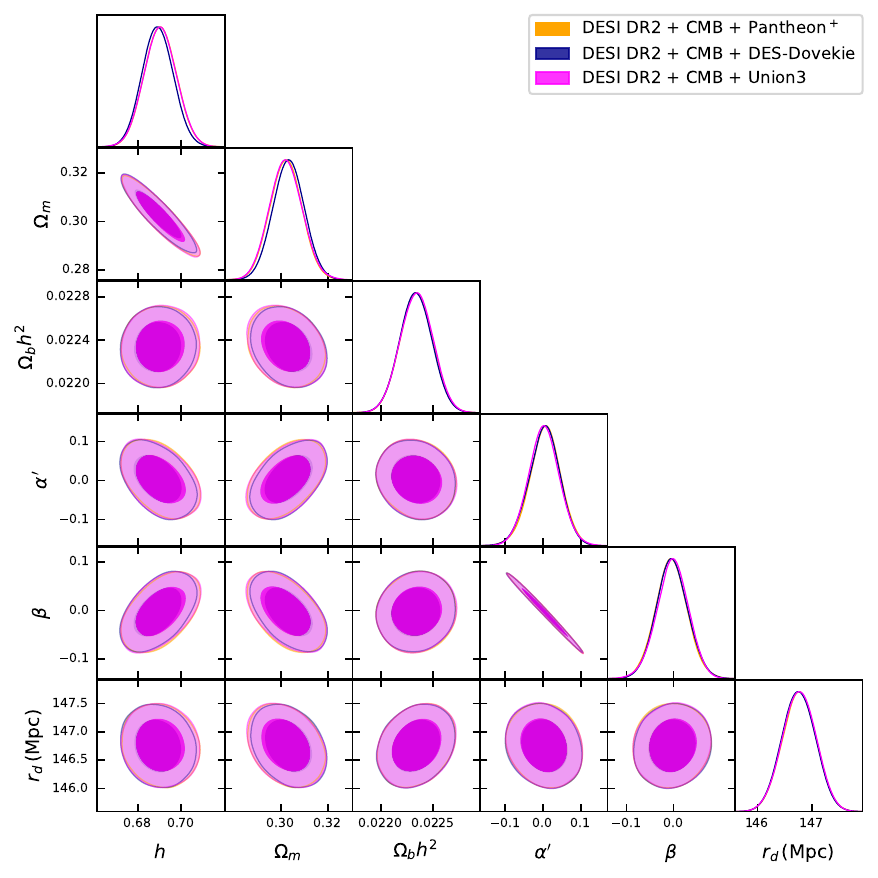}
    \label{fig_1a}
\end{subfigure}
\caption{This figure shows the confidence contours at the $1\sigma$ and $2\sigma$ levels for the parameters of the $\Lambda(t)$CDM model obtained using DESI DR2 BAO data combined with CMB shift parameters and different SNe Ia samples (Pantheon$^{+}$, DES-Dovekie, and Union3).}\label{fig_1}
\end{figure*}

Fig~\ref{fig_1} shows the corner plot for the parameters of the $\Lambda(t)$CDM model, obtained using different combinations of DESI DR2 with CMB and Type Ia supernova samples (Pantheon$^+$, DES-Dovekie, and Union3). The corner plot represents the off-diagonal plots as 2D contour plots, showing the correlations between different parameter pairs, while the diagonal terms show the 1D marginalized distributions. It is important to note that there is a negative correlation between the $\beta-\alpha$ Table~\ref{tab_1} presents the mean values along with the 68\% 1$\sigma$ confidence interval for the $\Lambda(t)$CDM model with different dataset combinations.

One can observe that in all dataset combinations, the $\Lambda(t)$CDM model predicts a higher value of $h$ compared to $\Lambda$CDM. Specifically, the $\Lambda(t)$CDM model shows a deviation of about $1.74\sigma$ for CMB+DESI DR2+Pantheon$^+$, $1.24\sigma$ for CMB+DESI DR2+DES-Dovekie, and $2.11\sigma$ for CMB+DESI DR2+Union3. The matter density parameter in the $\Lambda(t)$CDM model shows deviations from the $\Lambda$CDM prediction at the level of less than $0.6\sigma$, depending on the choice of the dataset combination. Specifically, the deviations are $0.51\sigma$ for CMB+DESI DR2+Pantheon$^+$, $0.38\sigma$ for CMB+DESI DR2+DES-Dovekie, and $0.51\sigma$ for CMB+DESI DR2+Union3.

Similarly, the baryonic density parameter $\Omega_b h^2$ exhibits only small deviations between the two models, with differences of $0.60\sigma$ for CMB+DESI DR2+Pantheon$^+$, $0.50\sigma$ for CMB+DESI DR2+DES-Dovekie, and $0.60\sigma$ for CMB+DESI DR2+Union3. In the case of the sound horizon, $r_d$, the deviations are slightly larger, with differences of $0.89\sigma$ for CMB+DESI DR2+Pantheon$^+$, $1.04\sigma$ for CMB+DESI DR2+DES-Dovekie, and $0.89\sigma$ for CMB+DESI DR2+Union3. These results shows that the $\Lambda(t)$CDM model cannot solve the Hubble tension, since an important aspect of the tension is that the combination of CMB + DESI~DR2 + Pantheon$^+$ must produce a higher value of $h$ together with a lower value of $r_d$, so that it can match $h = 0.735 \pm 0.014$~\cite{riess2022comprehensive} and $r_d = 138.0 \pm 1.0$~Mpc.

\begin{figure}
\centering
\includegraphics[scale=0.56]{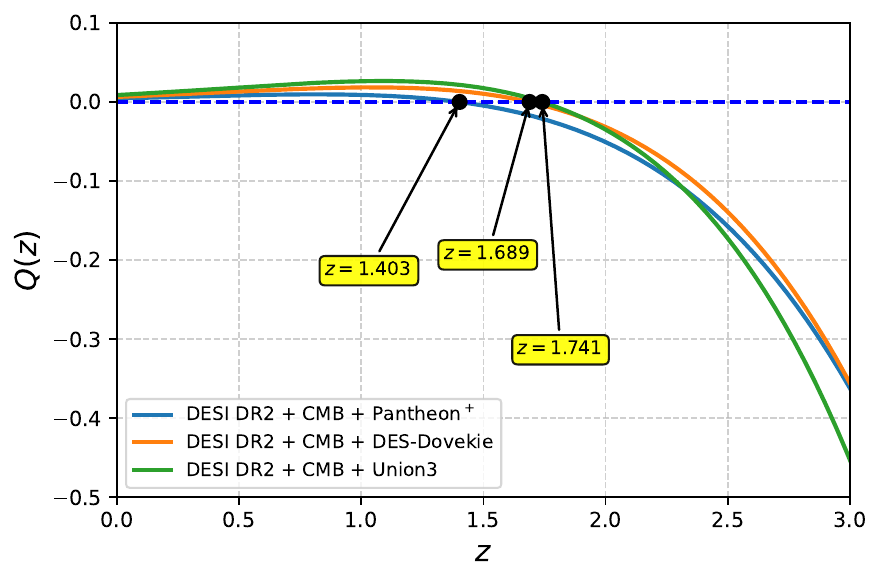}
\caption{This figure shows the evolution of the interaction term $Q(z)$ as a function of redshift for the $\Lambda(t)$CDM model using DESI DR2 BAO data combined with CMB shift parameters and different SNe Ia samples samples (Pantheon$^{+}$, DES-Dovekie, and Union3). The horizontal dashed line denotes $Q(z)=0$, shows the redshift at which the interaction changes sign for each dataset combination.}\label{fig_2}
\end{figure}

Fig.~\ref{fig_2} shows the redshift evolution of the interaction term $Q(z)$ for each dataset combination. In all cases, the interaction term is positive at low redshift, meaning vacuum decaying into dark matter. As the redshift increases, each dataset predicts a distinct transition redshift at which $Q(z)$ changes sign from positive to negative. Specifically, the zero crossing occurs at $z = 1.74$ for CMB+DESI+Pantheon$^+$, at $z = 1.68$ for CMB+DESI DR2+DES-Dovekie, and at $z = 1.40$ for CMB+DESI DR2+Union3. The negative interaction terms correspond to a decay of dark matter into the vacuum sector. This characteristic sign change is due to the fact that the $\Lambda(t)\mathrm{CDM}$ model predicts $\beta < 0$ and $\alpha' > 0$ in each combination.\\

The evolution of dark energy and dark matter can be understood through their respective solutions. Dark energy density evolves  approximately as
$ \rho_{de}(z)=(1+z)^{3(1+\omega)}\left[ \rho_{de,0}+ \int_0^z \frac{Q(z^\prime)}{H(z^\prime)(1+z^\prime)^{4+3\omega}}dz^\prime  \right].$
Clearly, for $Q=0$ standard evolution is recovered. Hence, the DE density evolves according to the sign of $Q(z)$. For $Q(z)>0$: The integral term is positive, but appears with a plus sign inside the bracket multiplied by $(1+z)^{3(1+\omega)}$. This causes $\rho_{de}$ to decay faster. At high redshift $z\geq 1$, DE is suppressed but not eliminated, allowing an early dark energy scenario. Also for $Q(z)<0$, the integral term becomes negative, effectively enhancing DE density at late times.

Similarly, DM density evolves approximately as $ \rho_{dm}(z)=(1+z)^{3}\left[ \rho_{dm,0}- \int_0^z \frac{Q(z^\prime)}{H(z^\prime)(1+z^\prime)^{4}}dz^\prime  \right].$
For $Q(z)>0$: The integral term is negative. Thus, matter density dilutes slower than $(1+z)^3$  results in enhancing early structure formation. Also for $Q(z)<0$: The integral term is positive. Matter density dilutes faster than $(1+z)^3$ results in suppression of structure growth. Statistically, the Bayesian evidence shows that the $\Lambda(t)$CDM model shows strong evidence over the $\Lambda$CDM model for all combinations of data sets, with $3 < |\Delta \ln \mathcal{Z}| < 5$.
\begin{table*}
\setlength{\tabcolsep}{5pt}
\resizebox{\textwidth}{!}{%
\begin{tabular}{lcccccccccc}
\hline
\textbf{Dataset / Model} & $h$ & $\Omega_m$ & $\Omega_b h^2$ & $\alpha'$ & $\beta$ & $r_d$(Mpc) & $|\Delta \ln \mathcal{Z}_{\Lambda\mathrm{CDM}, \mathrm{Model}}|$ \\
\hline
\textbf{$\Lambda$CDM} \\
DESI DR2 + CMB + Pantheon$^+$ & $0.680{\pm0.004}$ & $0.306{\pm0.005}$ & $0.02246 \pm 0.00013$ & --- & --- & $147.09 \pm 0.22$ & 0 \\
DESI DR2 + CMB + DES-Dovekie & $0.678{\pm0.004}$ & $0.307{\pm0.005}$ & $0.02251 \pm 0.00012$ & --- & --- & $147.13 \pm 0.23$ & 0 \\
DESI DR2 + CMB + Union3 & $0.680{\pm0.004}$ & $0.305{\pm0.005}$ & $0.02246 \pm 0.00013$ & --- & --- & $147.09 \pm 0.22$& 0 \\
\hline
\textbf{$\Lambda(t)$CDM} \\
DESI DR2 + CMB + Pantheon$^+$ & $0.689 \pm 0.007$ & $0.302 \pm 0.006$ & $0.02234 \pm 0.00016$ & $0.0051 \pm 0.047$ & $-0.0042 \pm 0.038$ & $146.76 \pm 0.30$ & 3.21 \\
DESI DR2 + CMB + DES-Dovekie & $0.688 \pm 0.007$ & $0.302 \pm 0.006$ & $0.02234 \pm 0.00016$ & $0.0038 \pm 0.046$ & $-0.0035 \pm 0.037$ & $146.75 \pm 0.30$ & 4.54 \\
DESI DR2 + CMB + Union3 & $0.690 \pm 0.007$ & $0.301 \pm 0.006$ & $0.02234 \pm 0.00015$ & $0.0069 \pm 0.045$ & $-0.0056 \pm 0.036$ & $146.76 \pm 0.31$  & 4.51 \\
\hline
\end{tabular}
}
\caption{This table shows the numerical values of the parameters obtained from the MCMC analysis for the $\Lambda(t)$CDM and $\Lambda$CDM models at the 68\% ($1\sigma$) confidence level, using DESI DR2 BAO data combined with CMB shift parameters and different SNe Ia samples (Pantheon$^{+}$, DES-Dovekie, and Union3).}\label{tab_1}
\end{table*}
\section{Dynamical system analysis}\label{sec_4}
\subsection{A short review of dynamical system analysis}
In mathematics, an ordinary differential equation (ODE) is a mathematical equation that consists of one independent variable with one or more dependent variable and their derivative with respect to the independent variable. An autonomous dynamical system is a system of ordinary differential equations where each differential equation of the system solely depends on the dependent variables and does not explicitly depend on the independent variable. Mathematically, a dynamical system can be expressed as
\begin{equation}\label{dyn}
    \frac{dX_i}{dt}=f_i\left(X_1\left(t\right),X_2\left(t\right),\dots,X_n\left(t\right)\right)
\end{equation}
Here $t$ is the independent variable and $f_i$ are n well-defined functions of the dependent variable $X_i$, $i=1,2,\dots,n$. Dynamical system analysis offers a powerful theoretical framework for investigating the qualitative dynamical behavior of complex physical models \citep{gonccalves2023cosmological,coley2003dynamical,bohmer2017dynamical}. For a particular choice of the independent variable $t=t_0$, the value of the dependent variable $X_i\left(t_0\right)$ is called the state of the system. The collection of all possible states is defined as the phase space of a dynamical system. If for any particular state, the system is in equilibrium, called the critical point or the equilibrium point. The critical points of the dynamical system \eqref{dyn} can be obtained by solving the system of nonlinear equations $f_i\left(X_i\left(t\right)\right)=0$. In the context of stability, a critical point can be classified as stable, unstable, or a saddle point, respectively. a critical point $X_i^*\left(t\right)$ of the dynamical system \eqref{dyn} will be considered to be a stable critical point if for any $\epsilon>0$, $\exists \hspace{0.1cm} \delta>0$ such that if $\xi_i\left(t\right)$ is a solution of the system \eqref{dyn} with $\lvert\lvert\xi_i\left(t_0\right)-X_i^*\left(t_0\right)\rvert\rvert<\delta$, then $\lvert\lvert\xi_i\left(t\right)-X_i^*\left(t\right)\rvert\rvert<\epsilon$, $\forall$ $t>t_0$. Similarly, a critical point $X_i^*\left(t\right)$ is asymptotically stable if for any $\epsilon>0$, $\exists$ $\delta>0$ such that if $\xi\left(t\right)$ be any solution of the dynamical system with $\lvert\lvert\xi_i\left(t_0\right)-X_i^*\left(t_0\right)\rvert\rvert<\delta$, then $\lim_{t\to\infty}\xi_i\left(t\right)=X_i^*\left(t\right)$. Mathematically, the linear stability theory can be used to determine the stability behavior of a critical point. In the linear stability theory, we can define the Jacobian matrix for the dynamical system \eqref{dyn} in the following way
\begin{center} $J
(X_1,X_2,\dots ,X_n)=\begin{pmatrix}
\frac{\partial f_1}{\partial X_1}&\frac{\partial f_1}{\partial X_2}&\dots&\frac{\partial f_1}{\partial X_n} \\
\frac{\partial f_2}{\partial X_1}&\frac{\partial f_2}{\partial X_2}&\dots&\frac{\partial f_2}{\partial X_n}\\
\vdots&\vdots&\vdots&\vdots\\
\frac{\partial f_n}{\partial X_1}&\frac{\partial f_n}{\partial X_2}&\dots&\frac{\partial f_n}{\partial X_n}\\
\end{pmatrix}$\\
\end{center}\par
In order to determine the stability criteria for a particular critical point $X_i^*$, first we compute the Jacobian matrix at this particular critical point as $J\left(X_1 ^*,X_2^*,\dots,X_n^*\right)$. Let $J_i^*$  are the eigenvalues of the Jacobian matrix $J\left(X_1 ^*,X_2^*,\dots,X_n^*\right)$ respectively. Then the stability of the critical point $X_i^*$ will depend on the signature of eigenvalues $J_i^*$ according to the following manner:
\begin{itemize}
    \item If the real part of all eigenvalues satisfies $\Re\left(J^\ast_i\right)>0,\forall i=1,2,\dots,n$, then $X^\ast_i$ is unstable.
    \end{itemize}
    \begin{itemize}
    \item If the real part of all eigenvalues satisfies $\Re\left(J^\ast_i\right)<0,\forall i=1,2,\dots,n$, then $X^\ast_i$ is stable.
    \end{itemize}
    \begin{itemize}
    \item If the eigenvalue spectrum contains both the positive and negative real part i.e. $\Re\left(J^\ast_i\right)<0$ and $\Re\left(J^\ast_j\right)>0$ for some  $i\neq j$, then $X^\ast_i$ is neither stable nor unstable, and it is called a saddle point.
    \end{itemize}
    If $J_i^*=0$ for some $i=1,2,\dots,n$, then the corresponding critical point $X_i^*$ is classified as a non-hyperbolic critical point and the linear stability theory fails to determine its stability behavior. In that scenario, some advanced methods of dynamical systems, like the center manifold theory or the Lyapunov functions, are useful to analyze the stability behavior. However, in the present manuscript, all the critical points are hyperbolic in nature, and the linear stability theory is sufficient to study the stability behavior. Due to the inherent nonlinearity of the field equation in cosmology, obtaining the exact analytical solution is challenging. However, by introducing suitable dynamical variables, these field equations can be reformulated as a nonlinear dynamical system. Within the framework of cosmology, the critical points of a dynamical system can be interpreted as representing distinct cosmological epochs \citep{gonccalves2023cosmological,coley2003dynamical,bohmer2017dynamical}. An ideal cosmological model is expected to exhibit critical points corresponding to important cosmological phases such as the inflation era, matter-dominated era, and late-time accelerated era. Generally, the critical point associated with inflation is unstable, while those corresponding to the matter-dominated era exhibit saddle-like behavior. In contrast, the critical point representing the late-time accelerated phase is generally stable, reflecting the asymptotic behavior of cosmological dynamics.
\subsection{Formulation of dynamical system for $\Lambda(t)$CDM Model}
In this section, we have explored a detailed dynamical behavior and corresponding cosmological evolution within the framework of $\Lambda(t)$CDM dark energy model by using the dynamical system theory. The dynamical system analysis method provides a rich mathematical technique for investigating highly nonlinear field equations. Moreover, this method allows us to investigate the evolution of cosmological solutions over time and offer a valuable insight into the nature of dark energy and late-time acceleration. 
We have considered the $\Lambda(t)$CDM model, where $\left(\alpha',\beta,\lambda_*\right)$ are constant parameter and $a=a(t)$ is the cosmic scale factor and $H=\frac{\dot{a}(t)}{a(t)}$ is the usual Hubble parameter.\\
To choose the dynamical variable, first ,we reformulate the first Friedmann equation in the following form 
\begin{equation}\label{eq2111}
    1=\left(\alpha' \frac{a^{-2}}{3H^2}+\frac{\beta}{3}+\frac{\lambda_*}{3H^2}\right)+\frac{8\pi G \rho_m}{3H^2}
\end{equation}
In order to formulate the dynamical system corresponding to $\Lambda(t)$CDM model, we have employed the following dynamical variable\\
\begin{equation}
    x=\frac{a^{-2}}{3H^2},y=\frac{1}{3H^2},z=\frac{\rho_m}{3H^2}.
\end{equation}
The dynamical variables $x$ and $y$ are related to the dark energy sector, while the variable $z=\frac{\rho_m}{3H^2}$ represents the matter density corresponding to the dark matter sector.

From equation \eqref{eq2111}, the expression of dark energy density $\Omega_\Lambda$ can be written in terms of the dynamical variable as
\begin{equation}
    \Omega_\Lambda=\alpha'x+\frac{\beta}{3}+\lambda_*y=1-z\label{eq22}
\end{equation}
Using these dynamical variables, the first field equation can be transformed as 
\begin{equation}
    \alpha' x+\frac{\beta}{3}+\lambda_*y+z=1
\end{equation}
Now, differentiating the dynamical variable with respect to $N=\log a(t)$ and using the conservation equation, we get the dynamical system corresponding to the generalised $\Lambda(t)$CDM model, which is equivalent to the Friedmann equation \eqref{eq13}-\eqref{eq14} as 
\begin{eqnarray}
    \frac{dx}{dN}&=&-2x\left(1+\frac{\dot{H}}{H^2}\right)\\
    \frac{dy}{dN}&=&-2y \frac{\dot{H}}{H^2}\\
    \frac{dz}{dN}&=&\frac{Q}{3H^3}-3z-2z\frac{\dot{H}}{H^2}
\end{eqnarray}
Now, due to the existence of the expression $\frac{\dot{H}}{H^2}$, the above dynamical system is not closed. From the second field equation \eqref{eq14}, we can write down the expression $\frac{\dot{H}}{H^2}$ in terms of the dynamical variable in the following way 
\begin{equation}\label{eq8}
    \frac{\dot{H}}{H^2}=\alpha'x+\lambda_*y-\frac{1}{2}z+\frac{\beta}{3}-1
\end{equation}
One can note that from equation \eqref{eq22}, the dynamical variable $z$ is dependent on $x$ and $y$. Therefore, we can eliminate the variable $z$ by using \eqref{eq22}, and in result we get a two dimensional dynamical system corresponding to the variable $x$ and $y$. Finally, by using equation \eqref{eq8} and eliminating $z$, we get the autonomous dynamical system as 
\begin{eqnarray}
    \frac{dx}{dN}&=&-x \left(\beta +3 \alpha'  x+3 \lambda_*  y-1\right) \label{eq9}\\
    \frac{dy}{dN}&=&-y \left(\beta +3 \alpha'  x+3 \lambda_*  y-3\right)\label{eq10}
\end{eqnarray}

Using equation \eqref{eq8}, the expression of effective EoS parameter $\omega_{eff}$ and deceleration parameter $q$ can be written in terms of the dynamical variables as 
\begin{eqnarray}\label{eq010}
    \omega_{eff}&=&-1-\frac{2}{3}\frac{\dot{H}}{H^2}=-\frac{\beta }{3}-\alpha'  x-\lambda_*  y \\
     q&=&-1-\frac{\dot{H}}{H^2}=\frac{1}{2} \left(-\beta -3 \alpha'  x-3 \lambda_*  y+1\right)\label{eq0010}
\end{eqnarray}
To study critical points and analyze stability, equations \eqref{eq9}-\eqref{eq10} are further solved as follows.
\subsection{Critical points and their stability analysis}
\begin{figure*}[htp]
\begin{subfigure}{.33\textwidth}
\includegraphics[width=\linewidth]{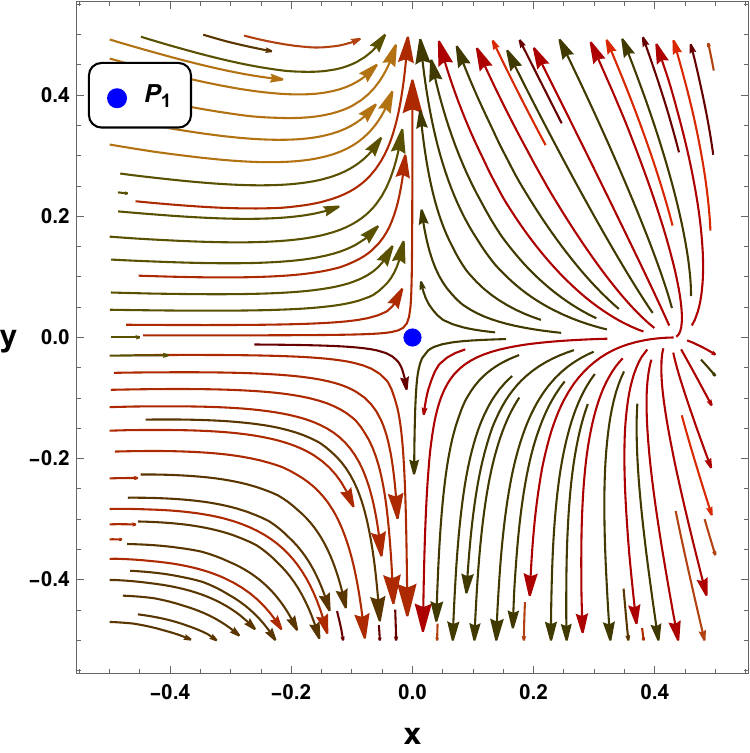}
    \caption{Phase space trajectories in $\left(x,y\right)$ plane corresponding to $P_1$}
    \label{fig_3a}
\end{subfigure}
\hfil
\begin{subfigure}{.33\textwidth}
\includegraphics[width=\linewidth]{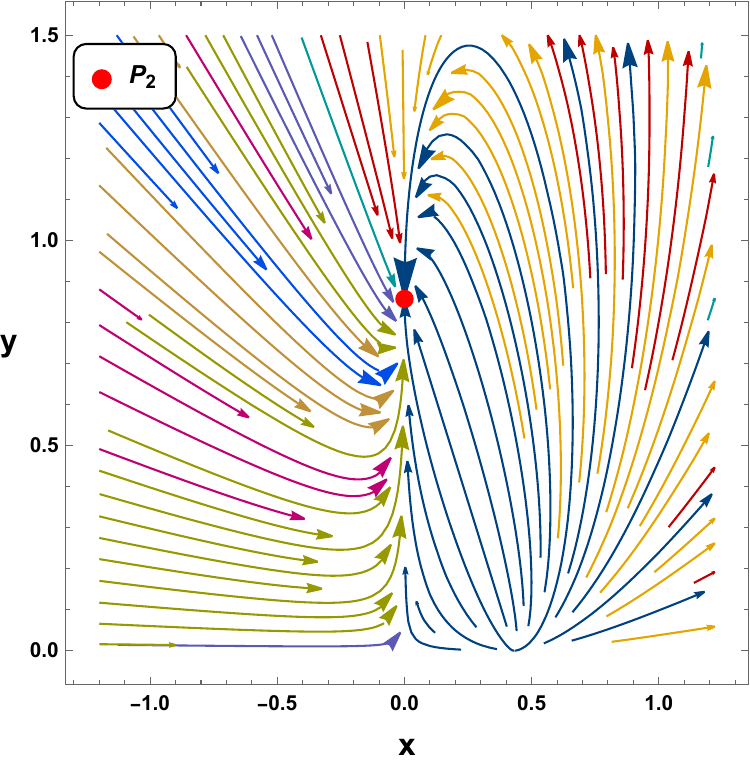}
    \caption{Phase space trajectories in $\left(x,y\right)$ plane corresponding to $P_2$}
    \label{fig_3b}
\end{subfigure}
\hfil
\begin{subfigure}{.33\textwidth}
\includegraphics[width=\linewidth]{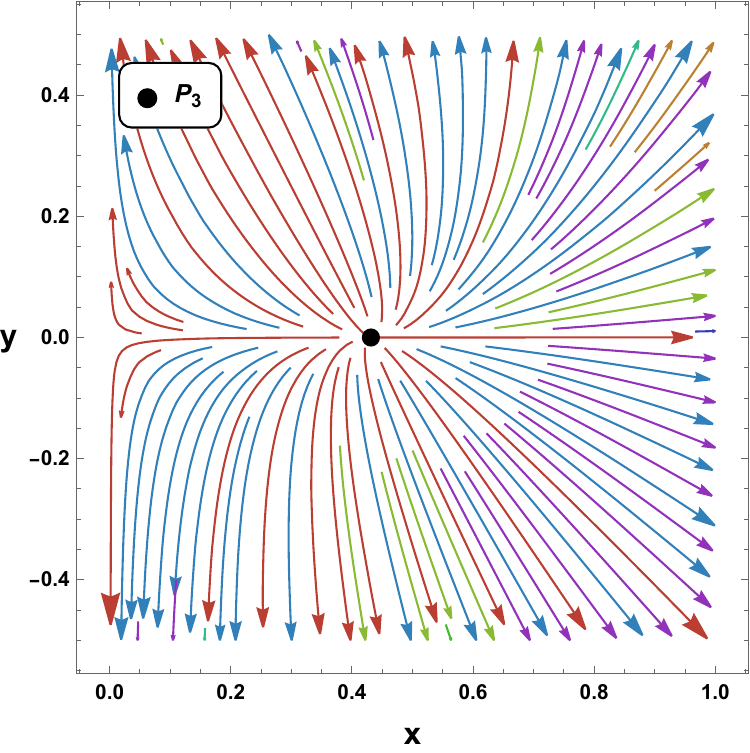}
    \caption{Phase space trajectories in $\left(x,y\right)$ plane corresponding to $P_3$}
    \label{fig_3c}
\end{subfigure}
\caption{Phase space trajectories in $\left(x,y\right)$ plane corresponding to the critical points for constrained parameter $\alpha'$ and $\beta$}\label{fig_3}
\end{figure*}

Here we investigate the physical characteristics and dynamical stability of the $\Lambda(t)$CDM cosmological model through the critical points of the dynamical system of equations \eqref{eq9}-\eqref{eq10}. To find the critical points of the dynamical system, we solve the system of nonlinear equations by setting $\frac{dx}{dN}=0$ and $\frac{dy}{dN}=0$, from equations \eqref{eq9}-\eqref{eq10} respectively. Each critical point is then associated with a specific epoch in the cosmological timeline by evaluating the corresponding energy density $\left(\Omega_m,\Omega_\Lambda\right)$, equation of state parameter $\omega_{eff}$, and the deceleration parameter $q$ respectively. The cosmological solution associated with the critical point corresponds to a fully dark energy-dominated universe when $\Omega_{\Lambda}=1$ and $\Omega_{m}=0$.Likewise, a matter-dominated universe can be characterized by $\Omega_m=1$ and $\Omega_\Lambda=0$. In addition, evaluating the effective equation-of-state parameter at a given critical point plays an important role in determining the specific cosmological epoch associated with that critical point. For instance, if $\omega_{eff}=0$ at a critical point, then the corresponding solution represents a matter era, while $-1<\omega_{eff}<-\frac{1}{3}$ corresponds to the quintessence era and $\omega_{eff}=-1$ signifies the later time de-sitter era respectively. Moreover, the dynamical stability features corresponding to each critical points is examined by the Lyapnov Linear stability method, which entails computing the eigenvalues of the Jacobian matrix \citep{gonccalves2023cosmological,coley2003dynamical,bohmer2017dynamical}. For the present $\Lambda\left(t\right)$CDM model, we have found a total of three critical points, namely $P_1, P_2$, and  $P_3$ respectively. The coordinates of critical points, along with their valid existence condition, are presented in Table \ref{tab_2}. Moreover, during the stability analysis of critical points, the standard existence condition has been employed, ensuring that the critical points lie within the viable physical region, specially the matter density parameter $\Omega_m$ remains non-negative and satisfies the condition $0\leq\Omega_m\leq1$. Consequently, the physically admissible two-dimensional phase space is 
\begin{equation*}
    \mathcal{S}=\left\lbrace \left(x,y\right)\in \mathbb{R}^2 : 0 \leq1-\alpha'x-\frac{\beta}{3}-\lambda_*y \leq1\right\rbrace
\end{equation*}
The trajectories in the phase space corresponding to the critical points is presented in Fig \ref{fig_3}.
A detailed analysis of dynamical stability with the physical characteristics in the cosmological context for each critical point is given in the next paragraph.
\begin{table*}
\centering
\begin{tabular}{|c|c|c|c|c|c|c|c|}
\hline
&&&&&&&\\
         Critical point &$x$&$y$&$\Omega_m$&$\Omega_\Lambda$&$\omega_{eff}$&$q$&Existence  \\
			&&&&&&&Condition\\
			\hline
			\hline
            &&&&&&&\\
$P_1$&$0$&$0$&$1-\frac{\beta }{3}$&$\frac{\beta}{3}$&$-\frac{\beta}{3}$&$\frac{1}{2}\left(1-\beta\right)$&$0\leq\beta\leq3$\\
&&&&&&&\\
\hline
&&&&&&&\\
			$P_2$&$0$&$-\frac{-3+\beta}{3\lambda_*}$&$0$&$1$&$-1$&$-1$&$\lambda_*\neq0$ \\
            &&&&&&&\\
            \hline
			&&&&&&&\\
            $P_3$&$-\frac{-1+\beta}{3\alpha'}$&$0$&$\frac{2}{3}$&$\frac{1}{3}$&$-\frac{1}{3}$&$0$&$\alpha'\neq0$\\
            &&&&&&&\\
\hline
\end{tabular}
\caption{Critical points along with their existence condition for $\Lambda(t)=\alpha' a^{-2}+\beta H^2+\lambda_{*}$}
\label{tab_2}
\end{table*}

\begin{itemize}
    \item \textbf{Critical point $P_1$:} The critical point $P_1$ consistently exists at the origin of the phase space. The corresponding value of matter and dark energy density parameters at this critical points are $\Omega_m=1-\frac{\beta}{3}$ and $\Omega_\Lambda=\frac{\beta}{3}$, respectively. As these parameters are functions of the model parameter $\beta$, varying the model parameter $\beta$ allows the system to emulate different cosmological epochs. For instance, setting $\beta=3$ yields $\Omega_m=0$ and $\Omega_\Lambda=1$, corresponding to a universe that is completely dominated by the dark energy sector. Conversely,  $\beta=0$, leads to $\Omega_m=1$ and $\Omega_\Lambda=0$, exhibits a matter dominated epoch. The effective EoS parameter is also $\beta$ dependent and it takes the form $\omega_{eff}=-\frac{\beta}{3}$. Thus, for $\beta=3$, one obtain $\omega_{eff}=-1$, representing de-sitter universe. For $\beta=0$, we get $\omega_{eff}=0$, which characterize the matter dominated phase. Furthermore, in the range $1<\beta<3$, the effective EoS parameter lies in the range $-1<\omega_{eff}<-\frac{1}{3}$, signifying a quintessence-like regime. The deceleration parameter corresponding to critical point $P_1$ is obtained as $q=\frac{1}{2}\left(1-\beta\right)$. Thus, for $\beta>1$, the condition $q<0$ holds, indicating the accelerated cosmological expansion.  In order to analyze the dynamical stability behavior corresponding to the critical point $P_1$, the set of eigenvalues of the Jacobian matrix is
    \begin{equation*}
        \left\lbrace1-\beta,3-\beta\right\rbrace
    \end{equation*}
    Due to the existence of nonzero eigenvalues, these critical points are hyperbolic, and hence their stability behavior can be determined by the linear stability theory. The stability features of the critical point $P_1$ are solely determined by the value of the model parameter $\beta$. Specifically, for $\beta>3$, both eigenvalues are negative, implying that the critical point is an attractor or stable node. In the range $1<\beta<3$, the eigenvalues are of opposite sign, and $P_1$ manifests as a saddle point. Finally, when $\beta<1$, both eigenvalues will be positive and the critical point will be an unstable node. The phase space trajectories in the neighborhood of $P_1$ is presented in Fig~\ref{fig_3a} corresponding to its saddle behavior.
\end{itemize}
\begin{itemize}
    \item \textbf{Critical point $P_2$:} The second critical point $P_2$ exists in the phase space under the specific condition on the model parameter $\lambda_*\neq0$. At this critical point, the energy density of the universe is entirely attributed to dark energy, with the density parameters taking the values $\Omega_m=0$ and $\Omega_\Lambda=1$ respectively. This configuration corresponds to a fully dark energy–dominated cosmological phase. The effective EoS parameter at this critical point remains fixed at $\omega_{eff}=-1$, indicating that the dark energy component behaves identically to a cosmological constant. Consequently, the cosmic expansion undergoes exponential acceleration, characterized by a constant negative deceleration parameter $q=-1$.This places the critical point $P_2$ firmly within the regime of de Sitter solutions. To study the stability behavior, the eigenvalues of the Jacobian matrix corresponding to $P_2$ are obtained as
    \begin{equation*}
        \left\lbrace-2,\beta -3\right\rbrace
    \end{equation*}
   Since both eigenvalues are non-zero, the point is classified as hyperbolic. For $\beta<3$,  both eigenvalues are negative, and therefore by the linear stability theory $P_2$ behaves as a stable attractor in the phase space. Otherwise, the critical point will exhibit a saddle nature. The phase space trajectories corresponding to the stable scenario of critical point $P_2$ is presented in Fig~\ref{fig_3b}.
\end{itemize}
\begin{itemize}
    \item \textbf{Critical point $P_3$:} The cosmological solution corresponding to the last critical point $P_3$ will represent a viable physical solution under the particular condition on the model parameter $\alpha'\neq0$. The critical point $P_3$ corresponds to a cosmological state in which both the matter and dark energy components coexist in non-negligible proportions. At this point, the matter and dark energy density parameters take the values $\Omega_m=\frac{2}{3}$ and $\Omega_{\Lambda}=\frac{1}{3}$, respectively. These combinations of energy density parameters indicate a transitional regime where the matter sector remains the dominant component, but the dark energy has already started to influence the cosmological dynamics. The effective EoS parameter associated with this critical point is $\omega_{eff}=-\frac{1}{3}$, which point out threshold between the decelarated and accelerated expansion era. Consistently, the deceleration parameter is found to be $q=0$, indicating a cosmological era, which is neither accelerating nor decelerating. This intermediate behavior makes $P_3$ an important point for understanding the transitional phase between the decelerated matter-dominated era to the dark energy-dominated accelerated expansion era. In order to study the stability behavior, the eigenvalues of the Jacobian matrix corresponding to $P_3$ are obtained as
    \begin{equation*}
        \left\lbrace2,-1+\beta\right\rbrace
    \end{equation*}
    Due to the existence of one positive eigenvalue, the stability of this critical point is not possible. Specifically, for $\beta<1$, the eigenvalues are of opposite sign and therefore, according to the linear stability theory, $P_3$ exhibits a saddle behavior. Otherwise, this critical point will behave as an unstable critical point. In Fig~\ref{fig_3c}, we have presented the phase space trajectories in the neighborhood of $P_3$ corresponding to its unstable/saddle characteristic.
\end{itemize}

\subsection{Cosmological evolution for $\Lambda(t)$CDM Model}
Following the analysis of the stability behavior and dynamical characteristics associated with each critical point, it is important to explore the asymptotic evolution of the underlying cosmological model.  In order to get a detailed understanding of the dynamical framework for the specific cosmological model, the evolution of the background cosmological parameters such as the matter density parameter $\Omega_m$, dark energy density parameter $\Omega_\Lambda$, effective EoS parameter $\omega_{eff}$, and deceleration parameter $q$ plays a significant role.

\begin{figure*}[htb]
\begin{subfigure}{.32\textwidth}
\includegraphics[width=\linewidth]{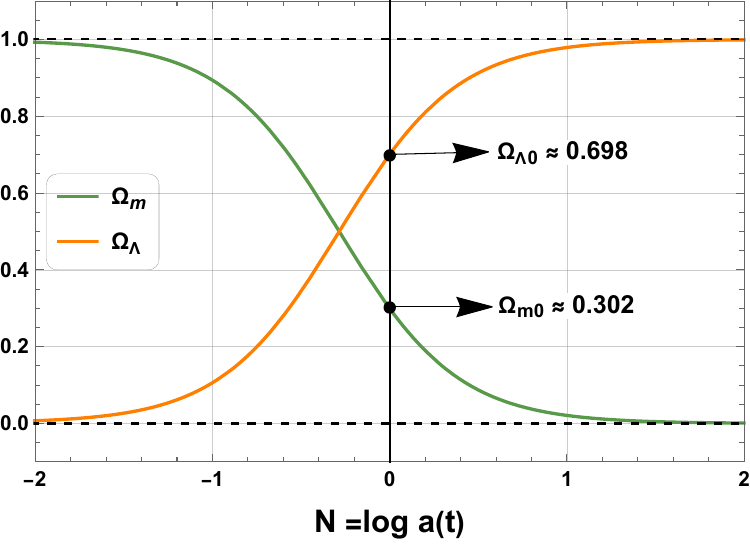}
\caption{}
    \label{fig_4a}
\end{subfigure}
\hfil
\begin{subfigure}{.32\textwidth}
\includegraphics[width=\linewidth]{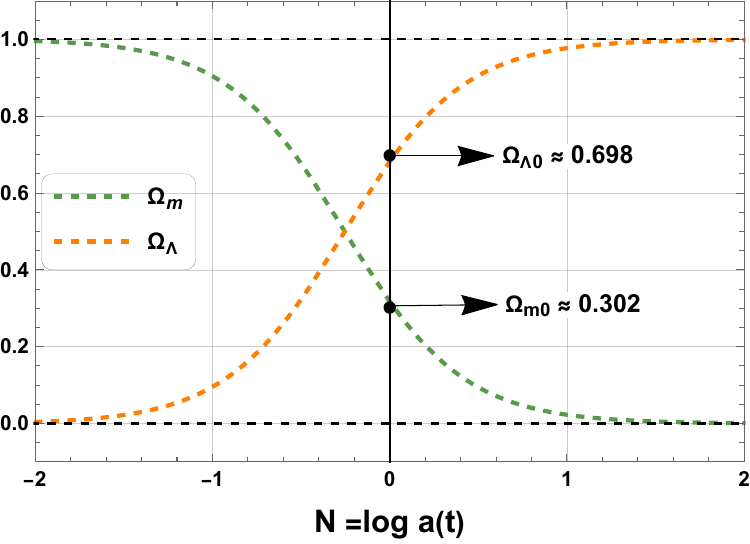}
\caption{}
    \label{fig_4b}
\end{subfigure}
\hfill
\begin{subfigure}{.32\textwidth}
\includegraphics[width=\linewidth]{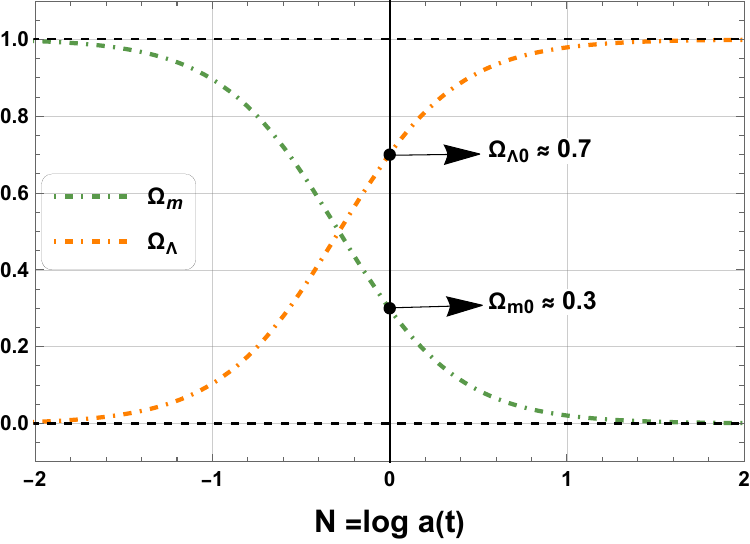}
\caption{}
    \label{fig_4c}
\end{subfigure}
\caption{This figure shows the numerical solutions of Eqs.~\eqref{eq22}, \eqref{eq010}, and \eqref{eq0010} for the evolution of the density parameters $\Omega_m$ and $\Omega_\Lambda$ as functions of $N \equiv \log a(t)$. The first, second, and third columns correspond to the dataset combinations DESI DR2 + CMB + Pantheon$^{+}$, DESI DR2 + CMB + DES-Dovekie, and DESI DR2 + CMB + Union3, respectively. The vertical line at $N=0$ denotes the present epoch, while $N<0$ and $N>0$ represent the past and future cosmic evolution, respectively.}
\label{fig_4}
\end{figure*}

\begin{figure*}
\begin{subfigure}{.32\textwidth}
\includegraphics[width=\linewidth]{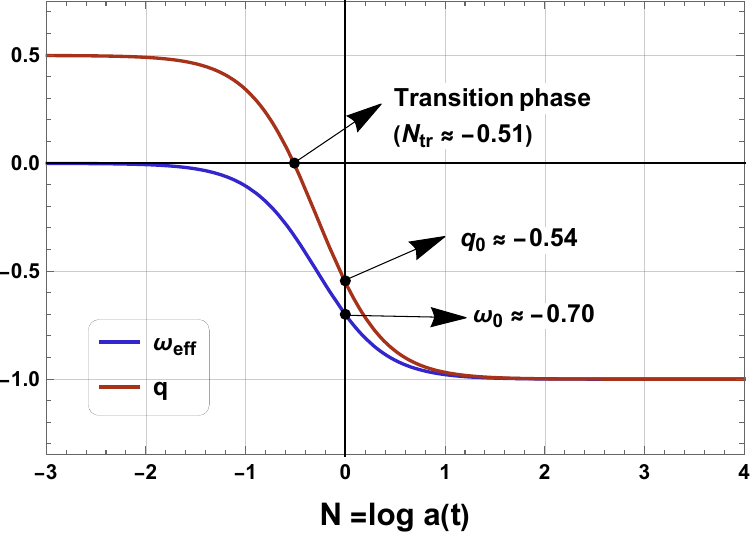}
\caption{}
    \label{fig5ax}
\end{subfigure}
\hfill
\begin{subfigure}{.32\textwidth}
\includegraphics[width=\linewidth]{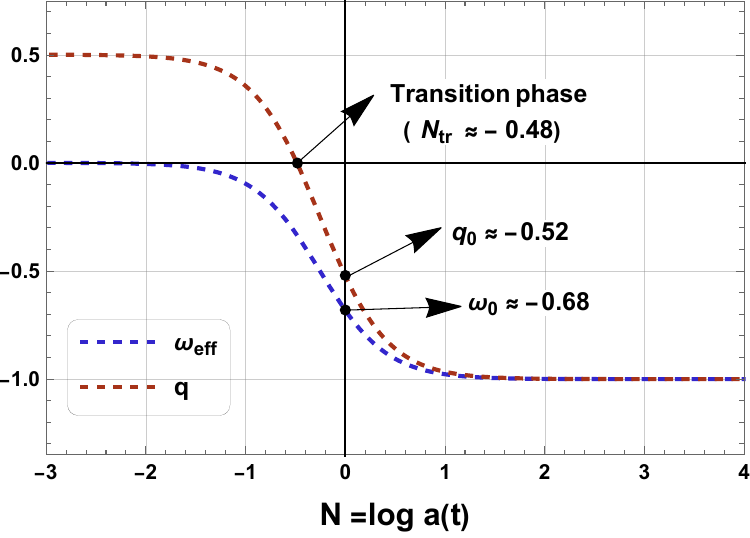}
\caption{}
    \label{fig5bx}
\end{subfigure}
\hfill
\begin{subfigure}{.32\textwidth}
\includegraphics[width=\linewidth]{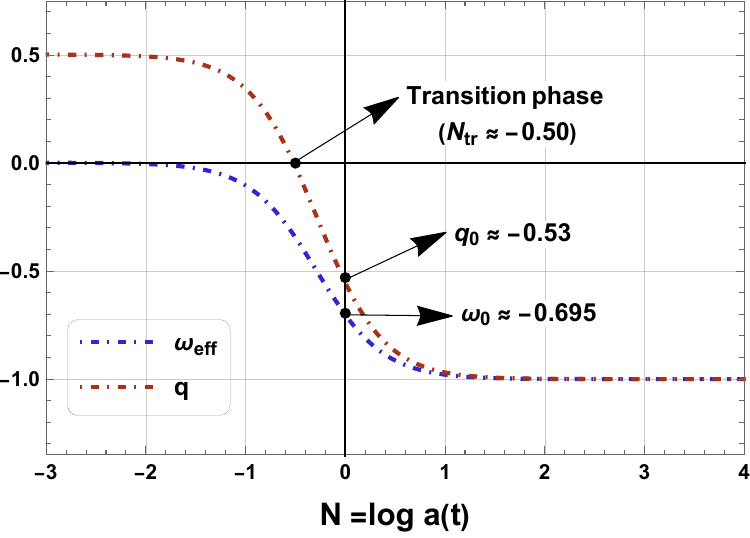}
\caption{}
    \label{fig5cx}
\end{subfigure}
\caption{This figure shows the numerical solutions of Eqs.~\eqref{eq22}, \eqref{eq010}, and \eqref{eq0010} for the effective EoS parameter $\omega_{\mathrm{eff}}$ and the deceleration parameter $q$ as functions of $N \equiv \log a(t)$. The first, second, and third columns correspond to the dataset combinations DESI DR2 + CMB + Pantheon$^{+}$, DESI DR2 + CMB + DES-Dovekie, and DESI DR2 + CMB + Union3, respectively. The vertical line at $N=0$ represents the present epoch, with $N<0$ and $N>0$ indicating the past and future epochs, respectively.}
\label{fig_5}
\end{figure*}

Some recent astronomical observations indicate that the universe is spatially flat, with present-day values of the density parameters approximately given by $\Omega_{\Lambda0}\approx0.7$ and $\Omega_{m0}\approx0.3$, respectively \cite{SNLS:2005qlf,SDSS:2005xqv}. We have numerically integrated the dynamical system \eqref{eq9}-\eqref{eq10} using different combinations of observational datasets with fine-tuned initial conditions to obtain a detailed description of the universe's evolution across several cosmological epochs. The numerical solution describing the evolution of the background density parameters $\Omega_m,\Omega_\Lambda$ are presented in Fig~\ref{fig_4}. Consequently, the evolution of effective EoS parameter $\omega_{eff}$ and the deceleration parameter $q$ is presented in Fig~\ref{fig_5}. The vertical line at $N = 0$ in these diagrams corresponds to the present cosmological epoch, with the region $N<0$ denoting the past epoch and the right half  $N>0$ representing the future epoch, respectively.

The evolution of the density parameters corresponding to different observational data combinations namely DESI DR2 + CMB + Pantheon$^{+}$, DESI DR2 + CMB + DES–Dovekie, and DESI DR2 + CMB + Union3 shows in Fig.~\ref{fig_4a}, Fig.~\ref{fig_4b}, and Fig.~\ref{fig_4c}, respectively. From the numerical solution of the dynamical system , the present-day value of the matter density parameter is found to be $\Omega_{m0}\simeq0.302$ for both DESI DR2 + CMB + Pantheon$^{+}$ and DESI DR2 + CMB + DES–Dovekie datasets, while for DESI DR2 + CMB + Union3 it is obtained as $\Omega_{m0}\simeq0.30$. Correspondingly, the present value of the dark energy density parameter is estimated as $\Omega_{\Lambda0}\simeq0.698$ for DESI DR2 + CMB + Pantheon$^{+}$, $\Omega_{\Lambda0}\simeq0.698$ for DESI DR2 + CMB + DES–Dovekie, and $\Omega_{\Lambda0}\simeq0.70$ for DESI DR2 + CMB + Union3, which are align to the results obtained from observational data analysis. Another important characteristic of the background density parameters can be revealed from \ref{fig_4}, that during the early Universe, the matter density significantly dominated the dark energy sector. However, as time progresses, the matter density $\Omega_m$ gradually decreases and the dark energy density $\Omega_{\Lambda}$ increases uniformly. Also, the evolution of density parameters shows that in the future, the matter density will be completely dominated by the dark energy density, i.e, $\Omega_\Lambda\to 1$ and $\Omega_m\to 0$, which indicates a dark energy-dominated Universe in late time.

The evolution of the effective EoS parameter $\omega_{eff}$ and the deceleration parameter $q$ for the different datasets are presented in Fig~\ref{fig_5}. One can note that from Fig~\ref{fig5ax}, Fig~\ref{fig5bx}, and Fig~\ref{fig5cx} the current value of the effective EoS parameter are obtained as $\omega_{eff}=-0.70$ (DESI DR2 + CMB + Pantheon$^{+}$), $\omega_{eff}=-0.68$ (DESI DR2 + CMB + DES–Dovekie) and $\omega_{eff}=-0.695$ (DESI DR2 + CMB + Union3) respectively.  Finally, the evolution of the deceleration parameter $q$ clearly shows the transition from the early decelerated era $\left(q>0\right)$ to the late-time accelerated expansion era $\left(q<0\right)$. The transition from decelerated era to the accelerated era for $\Lambda(t)$CDM cosmological model is obtained at $N_{tr}=-0.51$ (DESI DR2 + CMB + Pantheon$^{+}$), $N_{tr}=-0.48$ (DESI DR2 + CMB + DES–Dovekie) and $N_{tr}=-0.50$ (DESI DR2 + CMB + Union3) respectively. Also current value of the deceleration parameter for $\Lambda(t)$CDM model  corresponding to three different datasets indicated in Fig.~\ref{fig_5} are negative , which satisfies the observational results and represents the current accelerating expansion of the Universe.

\section{Conclusion}\label{sec_5}
In this work, we have performed a comprehensive observational and statistical analysis of the interacting $\Lambda(t)$CDM cosmological model using the recent DESI DR2 BAO measurements in combination with CMB and Type Ia supernova datasets, including Pantheon$^{+}$, DES-Dovekie, and Union3. The results have been systematically compared with the standard $\Lambda$CDM framework in order to assess the physical implications and statistical viability of vacuum dark matter interactions.

We find that the $\Lambda(t)$CDM model consistently predicts a slightly higher value of the Hubble parameter $h$ compared to $\Lambda$CDM across all dataset combinations, with deviations ranging from $1.24\sigma$ to $2.11\sigma$. However, the matter density parameter $\Omega_m$ and the baryon density $\Omega_b h^2$ remain fully consistent with the values predicted by the $\Lambda$CDM model, showing deviations well below the $1\sigma$ level. The sound horizon $r_d$ shows slightly lower values than those predicted by $\Lambda$CDM, and these are still insufficient to simultaneously increase $h$ and decrease $r_d$ enough to fully resolve the Hubble tension

The redshift evolution of the interaction term $Q(z)$ shows that for all dataset combinations the interaction is positive at low redshift, indicating a decay of the vacuum sector into dark matter, while at higher redshift it changes sign, indicating a decay of dark matter into the vacuum sector. The corresponding transition redshifts depend on the dataset combination and lie in the range $1.40 \lesssim z \lesssim 1.74$; this feature arises from the predicted model parameters $\beta < 0$ and $\alpha' > 0$. From a statistical perspective, the Bayesian evidence strongly favors the $\Lambda(t)$CDM model over the standard $\Lambda$CDM cosmology for all dataset combinations considered.

The dynamics of the cosmological models are studied through a dynamical system analysis. By formulating the field equations of the proposed cosmological model, we introduced appropriate dynamical variables to establish the corresponding nonlinear dynamical system. An ideal cosmological framework is expected to reproduce a sequence of evolutionary phases such as Inflation $\to$ matter dominated era $\to$ late-time accleration era. A critical point (or fixed point) in the phase space often corresponds to a cosmological regime (Matter domination, Dark energy domination (de Sitter phase) with constant negative EoS, or evolving dark energy with non-constant EoS for the constrained set of model parameter values. The dynamical system analysis revealed the existence of three critical points. The physical properties and the stability criteria of the critical points are highly sensitive to the parameters $\alpha',\beta$ and $\lambda_*$. In the range $1<\beta<3$, the critical point $P_1$ represents the dark energy-dominated accelerated quintessence era . Trajectories around the critical point $P_1$ are observed to have a saddle point.

The existence of such a critical point is cosmologically viable, since it represents the accelerated expansion phase for constrained values of $\beta$ and satisfies the required energy density conditions with a quintessence-like nature. The second critical point $P_2$ represents a complete dark energy dominated accelerated cosmological solution with constant EoS parameter value $\omega_{eff}=-1$ and $q=-1$. In terms of stability, this critical point is an attractor in the range $\beta<3$; this feature corresponds to the stable behavior and represents the de-sitter phase. Point $P_3$ has an unstable nature and represents the matter domination phase. Moreover, the value of the deceleration parameter at $P_3$ is obtained as $q=0$, indicates the transitional epoch from matter domination to the late time dark energy domination phase.  This outcome provides both theoretical backing and observational consistency for the proposed framework. Our analysis, based on the observational data combinations DESI DR2 + CMB + Pantheon$^{+}$, DESI DR2 + CMB + DES-Dovekie, and DESI DR2 + CMB + Union3, yields $\Omega_{\Lambda} \simeq 0.698$-$0.70$, an effective equation-of-state parameter $\omega_{\text{eff}} \simeq -0.68$ to $-0.70$, and a negative present-day deceleration parameter $q_0$, collectively indicating a phase of accelerated cosmic expansion driven by a dynamical dark energy component, distinct from the canonical $\Lambda$CDM scenario with $\omega = -1$.

Recent results from DESI DR2 provide new insights into the nature of dark energy, reinforcing the relevance of cosmological models beyond the $\Lambda$CDM model, such as $\Lambda(t)$CDM. In the present analysis, as $\Lambda(t)$CDM affects only the late-time expansion history, we primarily focus on the geometrical effects on the CMB rather than the full CMB spectrum. Moreover, the full CMB power spectra exhibit several mild anomalies that are not obviously geometrical in origin, such as the lensing anomaly and the lack of power on large angular scales, which may partially arise from residual systematics. These features can influence the inferred values of parameters of the $\Lambda(t)$CDM model and may spuriously suggest non-standard dark energy behavior even when none is physically present. A notable example is the apparent $(>2\sigma)$ preference for phantom dark energy from Planck data alone~\cite{escamilla2024state}.

Nevertheless, a more comprehensive analysis remains an important direction for future work. In particular, we plan to incorporate the full CMB power spectra, extend our framework to include the $\Lambda(t)$CDM model, and investigate its implications in greater depth. Moreover, we intend to study the impact of these scenarios on large-scale structure formation, making use of N-body simulations implemented with the \texttt{Gadget} code. These developments will allow for a more complete and robust assessment of dynamical dark energy models in light of current and forthcoming cosmological data.

\section*{Acknowledgments}
RM is thankful to UGC, Govt. of India, for providing Senior Research Fellowship (NTA Ref. No.: 211610083890).

\section*{Conflict Of Interest statement}
The authors declare that they have no known competing financial interests or personal relationships that could have appeared to influence the work reported in this document.

\section*{Author Contributions Statement} 
H.C. and M.B. performed the MCMC analysis and parameter estimation. R.M. carried out the dynamical system analysis and stability investigation. V.K.S. developed the original text and performed data organization and figure preparation. U.D. supervised the project, provided conceptual guidance, and reviewed the overall analysis. All authors contributed to discussions and reviewed the final manuscript.

\section*{Data Availability Statement} 
The datasets used in this manuscript are publicly available, and the code and numerical files will be made available upon reasonable request.

\bibliographystyle{cpc}
\bibliography{mybib.bib}

\end{document}